\documentclass[pdflatex,sn-mathphys-ay]{sn-jnl}

\usepackage{graphicx}%
\usepackage{multirow}%
\usepackage{amsmath,amssymb,amsfonts}%
\usepackage{amsthm}%
\usepackage{mathrsfs}%
\usepackage[title]{appendix}%
\usepackage{xcolor}%
\usepackage{textcomp}%
\usepackage{manyfoot}%
\usepackage{booktabs}%
\usepackage{algorithm}%
\usepackage{algorithmicx}%
\usepackage{algpseudocode}%
\usepackage{listings}%
\usepackage{braket}%
\usepackage{subcaption}%
\newtheorem{proposition}{Proposition}

\begin{document}

\title[Foreknowledge and Free Will Revisited]{Foreknowledge and Free Will Revisited}

\author*[1]{\fnm{Eliyahu Zvi} \sur{Engelberg}}\email{elien@jce.ac.il}

\affil*[1]
{
    \orgdiv{Department of Computer Science and Software Engineering}, 
    \orgname{Azrieli College of Engineering Jerusalem}, 
    \orgaddress
    {
        \street{26 Yaakov Shreibum Street}, \city{Jerusalem}, \postcode{9103501}, 
        \country{Israel}
    }
}

\abstract
{
    The apparent incompatibility between foreknowledge and free will is traditionally formulated
    in terms of the implication of knowledge of a future event for the freedom of that event. 
    This paper reexamines the problem using the notions of causal order and physical 
    determination provided by modern physics. 
    I consider measurements of two entangled particles at spacelike-separated events, 
    where a measurement outcome is taken to be free if it is not determined by the previous history 
    of the part of the universe accessible to it. 
    Relativity allows the temporal order of the two measurements to be reversed between reference 
    frames while preserving their causal independence. 
    Quantum correlations nevertheless allow an observer at one measurement event to know the outcome 
    at the other.
    I show that there is, therefore, 
    a frame in which an observer possesses foreknowledge of a measurement outcome that remains free 
    in this sense. 
    The argument demonstrates that epistemic access to a future event need not be grounded in physical 
    conditions that determine it.
}

\keywords{foreknowledge, free will, quantum entanglement, special relativity, causal order}



\maketitle

\section{Introduction} \label{sec:introduction}

If somebody possesses free will, can anybody know in advance how that person will act?
Let us assume that future contingency is possible, 
thus setting aside the logical fatalism raised in Aristotle's discussion of future contingents 
\citep{aristotle1984, swartz2004}.
Let us further assume that \emph{foreknowledge} of a future contingent is possible,
without attempting to explain its epistemic basis 
\citep{gettier1963, goldman1967, di_nucci2012, iacona2021, swartz2004}.
Even granting both assumptions, however, the traditional problem remains:
can an action be genuinely free if its occurrence is already known?
This question, 
raised by Cicero \citeyearpar{cicero1923} and subsequently developed by Boethius 
\citeyearpar{boethius1973},
continues to be discussed in classical and contemporary philosophy 
\citep{pike1965, van_inwagen1975, lewis1981, fischer1983, zagzebski1985, swartz2004}.

The word \emph{fore}knowledge suggests that this discussion is based on implicit assumptions about
the nature of time,
and possibly about the relationship between time and causality.
However, since the time in which the problem of foreknowledge and free will was first raised,
our understanding of time and causality has changed dramatically.
Specifically,
special relativity has shown that for spacelike-separated events,
the order of events is not absolute, but rather depends on the frame of reference.
What remains invariant is the causal ordering of events:
an event can influence another event only if the latter lies within the future light cone of the 
former \citep{rindler1982}.
Thus, temporal precedence and causal precedence,
which can easily be conflated in an intuitive discussion of foreknowledge,
must be distinguished \citep{zagzebski1985}.

The phrase \emph{free will}, as well, suggests some implicit assumptions.
In particular, it is normally assumed that free will, 
at least in the sense of libertarian free will, 
is incompatible with determinism \citep{pike1965, fischer1983}.
Quantum mechanics and its various interpretations, however,
have implications for determinism \citep{earman2004}.
Specifically,
the standard interpretation of quantum mechanics is indeterministic,
in the sense that the quantum state prior to measurement does not uniquely determine its outcome
\citep{born1962}.

It stands to reason, then, 
that the problem of foreknowledge and free will may benefit from considering the implications of 
modern physics in general, 
and of special relativity and quantum mechanics in particular.
This manuscript is an attempt to advance such a discussion.
Following Conway and Kochen \citeyearpar{conway2006},
I will say that a particle has ``free will'' if the outcome of a measurement on it is \emph{not} 
determined by the entire previous history of the part of the universe accessible to it.
Since the traditional problem arises from the apparent implication that foreknowledge renders a
future action necessary or determined \citep{pike1965, fischer1983, swartz2004},
this notion of free will suffices for the discussion.
In shorthand, I will say that such a particle is ``free'',
or that the measurement and its outcome are ``free''.

In the following sections, then, 
I will show that there are cases in which an observer can know the outcome of a measurement before 
it is made,
despite the fact that it is not determined at the time that the knowledge is acquired.
In other words, 
foreknowledge of the measurement outcome does not preclude its being free in the sense defined 
above.
I will do this by considering measurements on an entangled quantum state at two 
spacelike-separated events,
as observed in two different frames of reference.
I will restrict the discussion to the standard interpretation of quantum mechanics,
as described above.
I do not attempt to discuss the implications of deterministic interpretations of quantum mechanics,
such as Bohmian mechanics \citep{earman2004},
or to argue here for one interpretation over another.
The example I will present shows that epistemic access to a future event can be gained,
without that access being grounded in physical conditions that determine the event. 
The example therefore separates epistemic dependence from physical determination,
thus informing the discussion of foreknowledge and free will in the light of modern physics.

\section{Thought experiment setup} \label{sec:setup}

\begin{figure}[h]
    \centering
    \includegraphics[width=0.9\linewidth]{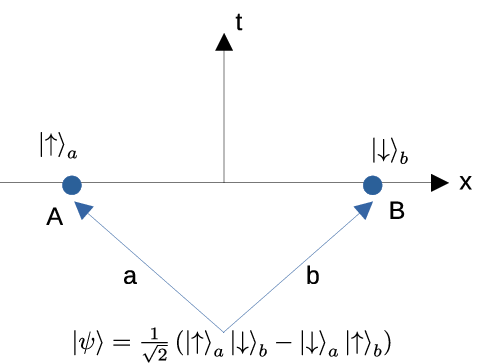}
    \caption
    {
        Two entangled electrons $a$ and $b$ are measured at two spacelike-separated events $A$ and 
        $B$.
    }
    \label{fig:setup:particles}
\end{figure}

Consider an entangled quantum system.
For definiteness of the example, 
let us consider a pair of entangled electrons $a$ and $b$ in the singlet state
\begin{equation}
    \ket{\psi} = \frac{1}{\sqrt{2}} 
    \left( \ket{\uparrow}_a \ket{\downarrow}_b - \ket{\downarrow}_a \ket{\uparrow}_b \right).
\end{equation}
The electrons $a$ and $b$ travel in opposite directions, 
and their spins are measured along the same axis at two spacelike-separated events in spacetime,
$A = (t_A = 0, x_A, y_A = 0, z_A = 0)$ and $B = (t_B = 0, x_B, y_B = 0, z_B = 0)$ respectively, 
see Fig.~\ref{fig:setup:particles}.
Here, $x_A < x_B$.
Going forward we will omit the $y$ and $z$ coordinates, so that $A = (0, x_A)$ and $B = (0, x_B)$.

Due to the measurements, the entangled state collapses into one of the two possible states,
$\ket{\psi} = \ket{\uparrow}_a \ket{\downarrow}_b$ or 
$\ket{\psi} = \ket{\downarrow}_a \ket{\uparrow}_b$.
Alice, who carried out the measurement at event $A$, 
will know the result of Bob's measurement at event $B$.
For example, if Alice measures $\ket{\uparrow}_a$, 
she will know that at the same time $t = 0$ Bob measured $\ket{\downarrow}_b$ \citep{bohm1957}.
Alice, then, has gained knowledge of the result of Bob's measurement.
In the standard interpretation of quantum mechanics,
the result of Bob's measurement is not determined until the time of measurement $t = 0$
\citep{born1962}.

\begin{figure}[h]
    \centering
    \includegraphics[width=0.9\linewidth]{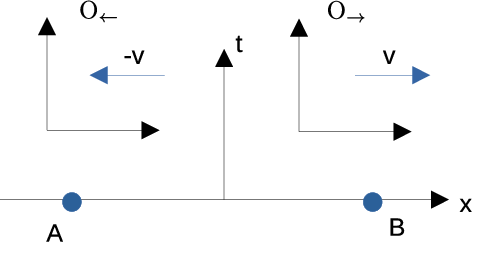}
    \caption
    {
        Two frames of reference, $\mathcal{O}_\leftarrow$ and $\mathcal{O}_\rightarrow$,
        are moving to the left and right respectively.
    }
    \label{fig:setup:frames}
\end{figure}

\begin{figure}[h]
    \centering
    \includegraphics[width=0.9\linewidth]{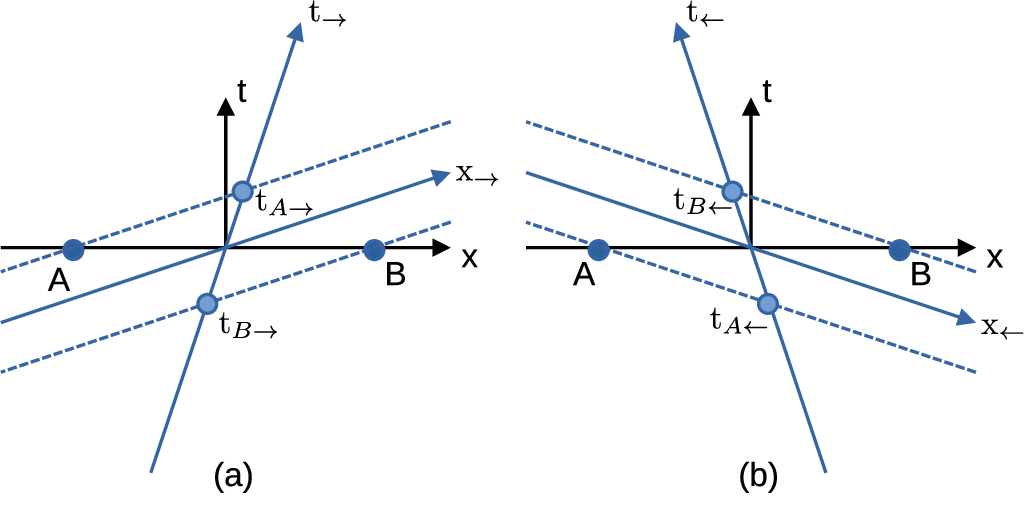}
    \caption
    {
        Spacetime diagrams of the events $A$ and $B$ in the frames of reference 
        (a) $\mathcal{O}_\rightarrow$, where $t_{A\rightarrow} > t_{B\rightarrow}$, and
        (b) $\mathcal{O}_\leftarrow$, where $t_{B\leftarrow} > t_{A\leftarrow}$.
    }
    \label{fig:setup:spacetime_diagrams}
\end{figure}

Now let us imagine that Alice and Bob are at rest.
Further, 
let us consider two additional frames of reference $\mathcal{O}_\leftarrow$ and 
$\mathcal{O}_\rightarrow$,
where $\mathcal{O}_\leftarrow$ is moving to the left and $\mathcal{O}_\rightarrow$ is moving to the 
right at a speed $v$ along the $x$ axis,
see Fig.~\ref{fig:setup:frames}.
We define the spacetime intervals $\Delta t = t_B - t_A$ and $\Delta x = x_B - x_A$.
The Lorentz transformations of the spacetime intervals are given by
\begin{equation}
    \begin{aligned}
        \Delta t' &= \gamma (\Delta t - v \Delta x) \\
        \Delta x' &= \gamma (\Delta x - v \Delta t)
    \end{aligned}
\end{equation}
where $\gamma = 1 / \sqrt{1 - v^2}$.
Here, the speed of light is set to 1, 
so that $v$ is a dimensionless number between 0 and 1 \citep{rindler1982}.
In the frame of reference $\mathcal{O}_\rightarrow$,
\begin{equation}
    \Delta t_\rightarrow = \gamma (\Delta t - v \Delta x) = \gamma v (x_A - x_B) < 0.
\end{equation}
Conversely, in the frame of reference $\mathcal{O}_\leftarrow$,
\begin{equation}
    \Delta t_\leftarrow = \gamma (\Delta t + v \Delta x) = \gamma v (x_B - x_A) > 0,
\end{equation}
see Fig.~\ref{fig:setup:spacetime_diagrams}.

In the frame of reference $\mathcal{O}_\rightarrow$, then, Bob's measurement precedes Alice's.
Thus, Alice's measurement cannot be a cause of Bob's measurement result:
although the two outcomes are correlated, Alice's measurement occurs later in this frame, 
while causal ordering is invariant under Lorentz transformations \citep{rindler1982}.
Moreover, in the standard interpretation of quantum mechanics, 
Bob's measurement outcome is not determined by the history accessible to electron $b$ prior to the 
measurement \citep{born1962}.
In other words, it is free in the sense defined above.
Since the causal past of Bob's measurement is invariant under Lorentz transformations 
\citep{rindler1982},
the measurement is free in all frames of reference.
This includes the frame of reference $\mathcal{O}_\leftarrow$,
where \emph{Alice's} measurement precedes \emph{Bob's}.
Therefore, in the frame of reference $\mathcal{O}_\leftarrow$,
Alice has foreknowledge of Bob's free measurement result.

\section{Formal Argument} \label{sec:formal_argument}

Before proceeding with the formal argument, I will introduce some notation.
First, to distinguish between logical implication and physical determination \citep{bradley1979},
I will use the symbol $\rightarrow$ to denote the former and $\leadsto$ to denote the latter.
For example, if proposition $R$ is ``It rained yesterday'' and proposition $W$ is
``The ground is wet today'',
then $R \rightarrow W$, $W \rightarrow R$, and $R \leadsto W$, but $W \not\leadsto R$.

Next, to denote knowledge, I will use the symbol $K$ \citep{hintikka1962}.
However, 
I want a notation that states not only that an agent knows the truth of a proposition or event,
but also \emph{when} the agent knows it.
So if an agent $X$ at a spacetime point $x^\mu$ knows that an event $e$ occurs,
I will write $K_{X,x^\mu} e$.
If an event $e_0$ occurs at the same spacetime point $x^\mu$,
I will allow myself to write $K_{X,e_0} e$.

Finally, following Belnap \citeyearpar{belnap1992}, 
I will use the symbol $<$ to denote \emph{causal order}.
For two events $e_1$ and $e_2$, 
I will write $e_1 < e_2$ if $e_1$ is in the past light cone of $e_2$.
Also following Belnap \citeyearpar{belnap1992}, I postulate transitivity of this relation:
\begin{proposition}
    If $e_1 < e_2$ and $e_2 < e_3$, then $e_1 < e_3$.
    \label{prop:transitivity_of_causal_order}
\end{proposition}
Unlike temporal order, the causal-order relation $<$ is Lorentz invariant: 
statements involving $<$ therefore hold independently of the choice of reference frame. 
By contrast, inequalities between time coordinates are frame dependent, 
and the relevant frame will be indicated explicitly by a subscript going forward.

According to the theory of relativity physical determination requires causal accessibility 
\citep{rindler1982},
so
\begin{proposition}
    If $e_1 \leadsto e_2$ then $e_1 < e_2$.
    \label{prop:causation_implies_causal_order}
\end{proposition}
The converse is not necessarily true,
because for a given event, not every event in its past light cone physically determines it.
Also according to the theory of relativity \citep{rindler1982},
\begin{proposition}
    There are three possible ways in which two distinct events can be related to each other:
    \begin{enumerate}
        \item $e_1 < e_2$, in which case $t_{e_1} < t_{e_2}$ in all frames of reference 
        (timelike or lightlike separation).
        \item $e_2 < e_1$, in which case $t_{e_2} < t_{e_1}$ in all frames of reference
        (timelike or lightlike separation).
        \item $e_1 \nleq e_2$ and $e_2 \nleq e_1$ (spacelike separation),
        in which case there exist frames of reference where $t_{e_1} < t_{e_2}$ and frames of 
        reference where $t_{e_2} < t_{e_1}$.
    \end{enumerate}
    \label{prop:causal_order_implies_time_order}
\end{proposition}
Here, $t_e$ is the time coordinate of event $e$ in a given frame of reference.

I will now show that Alice has foreknowledge of Bob's measurement result in the thought experiment 
described in Sec.~\ref{sec:setup}.
Let us denote Alice's measurement result as $S_a$,
and Bob's measurement result as $S_b$.
$S_a$ can assume the values $S = \uparrow$ or $S = \downarrow = \neg \uparrow$.
Alice knows that $S_b = \neg S_a$ because the spins of the electrons are entangled. 
When she makes her measurement and discovers that $S_a = S$,
she knows the value of $S_b$ as well:
\begin{equation}
    K_{\mathcal{A},A} \left(S_b = \neg S_a \;\wedge\; S_a = S\right)
    \qquad \Rightarrow \qquad
    K_{\mathcal{A},A} \left(S_b = \neg S\right).
    \label{eq:alice_knows_bobs_spin}
\end{equation}
Here, $\mathcal{A}$ denotes Alice as an agent.
Let $B$ denote the event in which Bob measures $S_b = \neg S$.
Assuming that Alice knows that Bob performs the specified measurement, 
Eq.~(\ref{eq:alice_knows_bobs_spin}) therefore gives
\begin{equation}
    K_{\mathcal{A},A} B.
    \label{eq:alice_knows_bobs_result}
\end{equation}
Since $A$ and $B$ in the thought experiment are spacelike separated,
according to Proposition \ref{prop:causal_order_implies_time_order} there is a frame of reference 
$\mathcal{O}$ in which $t_{\mathcal{O}A} < t_{\mathcal{O}B}$.
Together with Eq.~(\ref{eq:alice_knows_bobs_result}) we have
\begin{equation}
    \exists\, \mathcal{O}: \qquad
    t_{\mathcal{O}A} < t_{\mathcal{O}B} \;\wedge\; K_{\mathcal{A},A} B,
    \label{eq:alice_has_foreknowledge}
\end{equation}
so Alice has foreknowledge of Bob's measurement result.

Next, I will prove by contradiction that Bob's measurement is free.
Let us keep in mind that Alice and Bob's measurements are spacelike separated,
so
\begin{equation}
    A \nleq B \;\wedge\; B \nleq A
    \label{eq:alice_bob_spacelike_separated}
\end{equation}
(Proposition \ref{prop:causal_order_implies_time_order}).
Let us also remember that in the standard interpretation of quantum mechanics,
before either spin measurement of the entangled state has occurred, 
neither outcome is determined \citep{bohm1957}.
Consequently, if some event $e$ physically determines Bob's outcome, 
there can be no frame of reference in which $e$ precedes both $A$ and $B$.
Thus,
\begin{equation}
    \forall \mathcal{O}, \qquad
    \neg\left(t_{\mathcal{O}e} < t_{\mathcal{O}A} 
    \;\wedge\; 
    t_{\mathcal{O}e} < t_{\mathcal{O}B}\right).
    \label{eq:bell}
\end{equation}

\begin{figure}[h]
    \centering
    \includegraphics[width=0.9\linewidth]{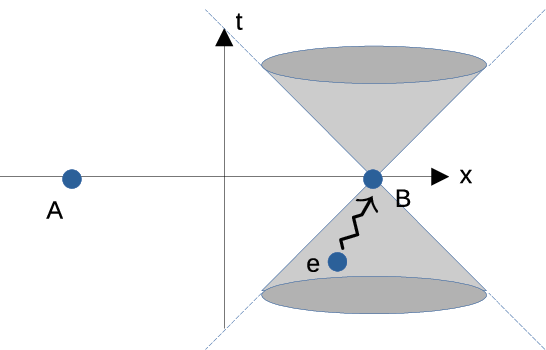}
    \caption
    {
        An event $e$ completes the physical determination of $B$,
        so that $e < B \;\wedge\; A \nleq B \;\wedge\; B \nleq A$.
        It is shown in the text that this is impossible if $A$ and $B$ are the measurement events
        of entangled spins.
    }
    \label{fig:formal:e_lesser_than_b}
\end{figure}

Now assume that Bob's measurement is not free,
in the sense that the physical history of the universe accessible to electron $b$,
prior to the measurement, determines its outcome.
Let $e$ denote the event at which this determination becomes complete,
so that
\begin{equation}
    e \leadsto B \qquad \text{(Assumed for contradiction)},
    \label{eq:assumption_bobs_measurement_not_free}
\end{equation}
see Fig.~\ref{fig:formal:e_lesser_than_b}.
According to Proposition \ref{prop:causation_implies_causal_order}, then,
\begin{equation}
    e < B \qquad \text{(Derived from assumption which will be contradicted)}.
    \label{eq:assumption_bobs_measurement_not_free_causal_order}
\end{equation}

What can be said about the causal order of $A$ and $e$?
Let us examine the three possibilities described in Proposition 
\ref{prop:causal_order_implies_time_order}:
\begin{enumerate}
    \item $A < e$ is impossible, because then $A < e < B$ according to Proposition 
    \ref{prop:transitivity_of_causal_order}, 
    but this contradicts Eq.~(\ref{eq:alice_bob_spacelike_separated}).
    \item $e < A$ is impossible, 
    because then $t_{\mathcal{O}e} < t_{\mathcal{O}A}$ and $t_{\mathcal{O}e} < t_{\mathcal{O}B}$ in 
    all frames of reference,
    due to Eq.~(\ref{eq:assumption_bobs_measurement_not_free_causal_order}) and Proposition
    \ref{prop:causal_order_implies_time_order},
    but this contradicts Eq.~(\ref{eq:bell}).
    \item $A \nleq e$ and $e \nleq A$ is impossible,
    because then $t_{\mathcal{O}e} < t_{\mathcal{O}B}$ in \emph{all} frames of reference due to 
    Eq.~(\ref{eq:assumption_bobs_measurement_not_free_causal_order})
    and Proposition \ref{prop:causal_order_implies_time_order},
    and $t_{\mathcal{O}e} < t_{\mathcal{O}A}$ in \emph{some} frames of reference due to the same 
    Proposition,
    but this again contradicts Eq.~(\ref{eq:bell}).
\end{enumerate}
Thus, all three possible causal orders of $A$ and $e$ lead to contradictions.
Therefore the assumption in Eq.~(\ref{eq:assumption_bobs_measurement_not_free}) is false,
and Bob's measurement is free,
even though Alice has foreknowledge of its result. \qed

\section{Discussion} \label{sec:discussion}

\subsection{The Free Will Theorem} \label{subsec:discussion:free_will_theorem}

The argument in this paper has some similarities to Conway and Kochen's \citeyearpar{conway2006} 
Free Will Theorem.
The Free Will Theorem connects the free choice of measurement settings by the experimenters to the 
freedom of the responses of the particles,
in the sense that their response cannot be determined by their respective prior accessible 
histories.
To establish this result, Conway and Kochen employ three axioms: 
SPIN constrains the possible responses of a spin-1 particle to measurements along orthogonal 
directions; 
TWIN establishes the appropriate correlations between the responses of the two entangled particles;
and FIN restricts causal influence in accordance with relativistic causal structure. 
The argument from SPIN relies on the Kochen-Specker \citeyearpar{kochen1967} result 
\citep{conway2006, cator2014}.

In contrast, the argument in this paper does not attempt to connect the free agency of the 
experimenters to the freedom of the responses of the particles.
Rather, 
it shows that the freedom of the response at event $B$ is compatible with foreknowledge of
an agent at event $A$.
Here, the important point is not that the agent is free, 
but rather that the agent can have knowledge of events, and specifically, in the present case, 
knowledge of a free event prior to its occurrence.
To illustrate the distinction between the premises of the two arguments,
imagine that Alice's choice of the measurement setting were not free,
but that Alice nevertheless acquired knowledge of the outcome at $B$.
Nothing in the present argument would be affected, because what matters at $A$ is Alice's knowledge,
rather than the freedom of the outcome she observes.
By contrast, 
the freedom of the experimenter's choice of setting is a necessary premise of the Free Will 
Theorem.
Conversely, suppose that Alice's choice of the setting were free, 
but that no agent at $A$ acquired knowledge of the outcome at B. 
This would not undermine the Free Will Theorem, 
but it would eliminate the foreknowledge required by the present argument.

Therefore, 
although the present argument adopts Conway and Kochen's definition of freedom, 
it does not rely on the Free Will Theorem itself. 
The freedom of the outcome at B, in the sense adopted here, 
is established independently in Secs.~\ref{sec:setup} and \ref{sec:formal_argument} using relativistic 
causal order, 
together with the assumption of standard quantum mechanics,
that the outcome of a measurement is not determined prior to the measurement \citep{born1962}.
Since in the case considered here two entangled measurements are made,
what should be considered the time of \emph{the} measurement, either $t_A$ or $t_B$, is ambiguous.
In Sec.~\ref{sec:formal_argument}, I allowed both likely interpretations of the time of measurement,
and required only that the event be undetermined \emph{either} before $t_A$ \emph{or} before $t_B$,
thus arriving at Eq.~(\ref{eq:bell}).
This sufficed for continuing the argument, and for establishing the freedom of the outcome at $B$.
The SPIN axiom and the associated Kochen-Specker result are therefore not required,
essentially because the thesis I wish to establish is different from that of the Free Will Theorem.

\subsection{The standard interpretation of quantum mechanics}

Another type of experiments that are similar to the thought experiment considered here are Bell-type
experiments,
which are designed to test for violations of Bell inequalities.
Bell-type experiments, in their textbook form, 
concern correlations between measurements performed with alternative settings at spacelike-separated 
locations.
Their derivation involves an independence assumption concerning the choice of measurement settings, 
commonly referred to as measurement independence or freedom of choice.
In particular, the settings chosen by the experimenters are assumed to be independent of the 
variables characterizing the measured systems.
Together with the appropriate locality assumptions, 
this leads to inequalities which are violated by the predictions of quantum mechanics 
\citep{bell1964, sakurai1994}.

Such an argument might appear to provide a shortcut to the conclusion of 
Sec.~\ref{sec:formal_argument} that Bob's measurement is free.
However, 
doing so would introduce an assumption concerning the freedom or independence of the experimenters' 
choices.
This is similar in one respect to the Free Will Theorem discussed above:
in both cases, 
assumptions concerning the experimenters' choices play a role in drawing conclusions about the 
physical systems being measured.
The present argument is constructed so as to avoid such an assumption.
The relevant property of the experimenters here is not the freedom of their choices,
but the knowledge they acquire from the measurement outcomes.

For this reason, the argument of Sec.~\ref{sec:formal_argument} begins instead from the standard 
indeterministic interpretation of quantum mechanics adopted in this work.
An individual measurement outcome is not determined prior to measurement \citep{born1962}.
In the entangled system considered here, however, 
measurement takes place at two spacelike-separated events, $A$ and $B$, 
for which there is no invariant temporal ordering.
Rather than assuming that Bob's outcome becomes determined specifically at either $A$ or $B$, 
I therefore impose only the weaker condition that its determination cannot already be complete 
before both measurement events.
Thus, if $e$ is an event sufficient to physically determine Bob's outcome, 
there exists no frame of reference in which $e$ precedes both $A$ and $B$.
This is the condition expressed, once again, by Eq.~(\ref{eq:bell}).
It allows, in particular, 
for the possibility that the determination of the outcome may in \emph{some} frames precede $A$ but not 
$B$,
and in \emph{other} frames precede $B$ but not $A$.

There is also an important difference between the physical setups involved in Bell-type experiments 
and in the present argument.
The violation of a Bell inequality requires consideration of different measurement settings at two 
events,
as well as alternative settings at each event,
because correlations between the outcomes associated with these incompatible measurements are 
essential to the violation \citep{bell1964, sakurai1994}.
In contrast, 
the present argument requires the experimenters to measure the \emph{same} observable at both events,
so that Alice's knowledge of Bob's outcome will be established from her own measurement result.
Consequently, 
no inference from the violation of a Bell inequality is required for the argument of 
Sec.~\ref{sec:formal_argument},
except insofar as such violations may be taken as evidence relevant to the choice among 
interpretations of quantum mechanics.
That, however, is outside the scope of this paper.
Bell inequalities and the present argument, then, concern related features of entangled systems,
but they examine different setups, employ different assumptions, and establish different conclusions.

\subsection{Determining events and causal histories}

In Sec.~\ref{sec:formal_argument}, 
physical determination of Bob's outcome was represented by a single event $e$,
defined as the event at which the determination becomes complete.
Strictly speaking, however,
the definition of freedom adopted in this paper concerns determination by the entire previous
accessible history,
rather than by a single event.
The relation between a complete causal history and the identification of individual determining 
events is itself nontrivial, 
and may depend on the choice and granularity of the causal model \citep{halpern2005}.

The present argument does not attempt to resolve this more general problem.
It assumes that, if the previous history physically determines an outcome,
then there is some event $e$ at which the conditions sufficient for that determination become
complete.
Whether the argument can be generalized to cases in which determination cannot be reduced to such an
event is an interesting question for future study.

\subsection{Complete foreknowledge}

In Sec.~\ref{sec:setup}, I wrote that Alice knows the result of Bob's measurement.
Alice definitely knows the result \emph{if} Bob makes the measurement,
but can she be sure that Bob will actually make it?

I stated in Sec.~\ref{subsec:discussion:free_will_theorem} that the argument of 
Secs.~\ref{sec:setup} and \ref{sec:formal_argument} does not require Alice's choice of the 
measurement setting to be free.
The same can be said about Bob's choice to perform the measurement, 
because the freedom established for his measurement outcome is independent of his own agency.
Both Alice's foreknowledge and Bob's free outcome exist, then,
even if Bob is preprogrammed to make his measurement,
and even if Alice knows that he is preprogrammed to do so.

One might argue, however, 
that Alice's foreknowledge cannot be complete because there may be external factors, 
other than Bob's free will, that prevent him from making the measurement.
These can be either factors within Bob's past light cone but outside Alice's,
or factors within Alice's light cone that are not known to her.
While this is true, it is an argument concerning the general possibility of complete foreknowledge,
regardless of the existence of free will.
Thus, insofar as foreknowledge is possible, 
the present argument shows that the presence of free will, 
understood as the absence of physical determination by the accessible past history, 
does not by itself preclude it.
This type of foreknowledge might, indeed, 
differ from the complete foreknowledge discussed in theological formulations of the problem
\citep{pike1965}.
More generally, however,
applying the present argument to the theological problem would require additional adaptations 
beyond the notion of foreknowledge considered here. 
Such an extension is beyond the scope of this paper.

\subsection{An epistemological approach to the problem}

The argument of Sec.~\ref{sec:formal_argument},
showing that a free event can be known in advance,
relies on an analysis of the causal order of events in spacetime.
This suggests an epistemological approach to the problem of foreknowledge and free will,
alongside the traditional modal and ontological formulations.
Such a problem might be formulated along the following lines:
\begin{enumerate}
    \item If $\mathcal{A}$ knows that $B$ will happen, 
    then this knowledge must be epistemically grounded in a set of events $\{e\}$ sufficient to 
    physically determine $B$.
    \item If $\mathcal{A}$'s knowledge of $B$ is grounded in $\{e\}$, 
    then $\{e\}$ must precede the event at which $\mathcal{A}$ knows $B$.
    \item If $\{e\}$ precedes $\mathcal{A}$'s knowledge of $B$, 
    and $\mathcal{A}$'s knowledge of $B$ precedes $B$, then $\{e\}$ precedes $B$.
    \item Therefore, a set of events $\{e\}$, sufficient to determine $B$, has happened before $B$.
    Thus $B$ is determined by past events, and is not free.
\end{enumerate}
In essence, one asks: 
\emph{what could epistemically ground knowledge of a later event,
if not its determining antecedents?}
While some modal formulations of the apparent conflict can be addressed by distinguishing the 
necessity of the consequence from the necessity of the consequent \citep{talbott1993, westphal2011}, 
the epistemological formulation considered here is unaffected by this distinction: 
even if $B$ remains contingent, 
there is still the question of what could epistemically ground knowledge of $B$.

The present argument provides a counterexample to the first step of this epistemological 
formulation.
In the thought experiment described in Sec.~\ref{sec:setup},
Alice's knowledge of $B$ is not grounded in conditions physically determining $B$,
but in knowledge of another event that logically implies $B$ without physically determining it.
Indeed, event $B$ is free in the sense adopted here, 
so it has no determining antecedents in its accessible past.
In this respect, knowledge of a spacelike-separated event resembles knowledge of a past event:
in both cases, 
knowledge of the event can be grounded otherwise than in knowledge of its determining antecedents.
This raises the possibility that the same epistemic relation could obtain when the event of 
knowledge lies in the past light cone of the known event, 
thereby constituting foreknowledge in the ordinary temporal sense. 
The present argument does not establish this possibility, 
but suggests it as a direction for further investigation.

\section{Conclusion} \label{sec:conclusion}

\begin{figure}[h]
    \centering
    \includegraphics[width=0.9\linewidth]{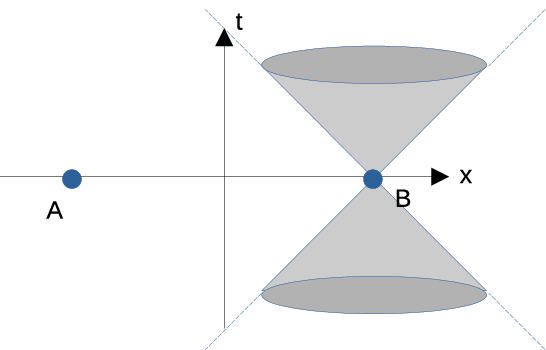}
    \caption
    {
        Alice's knowledge is acquired outside the light cone of Bob's measurement result.
    }
    \label{fig:conclusion:light_cone}
\end{figure}

I have described a case in which an agent can possess foreknowledge of a free event.
The crux of the argument depends on the fact that the event at which the knowledge is acquired is 
outside the light cone of the event \emph{about} which the knowledge is acquired,
see Fig.~\ref{fig:conclusion:light_cone},
so that the two events are not causally ordered.
Thus, under the definition of freedom adopted in this work,
foreknowledge and freedom are not intrinsically incompatible.

In a broader sense,
modern physics provides notions of precedence and causality that differ from classical intuitions.
We learn from relativity that precedence is not absolute:
for spacelike-separated events, their temporal order depends on the reference frame, 
whereas their lack of causal ordering is invariant.
On the other hand, 
within the framework of standard quantum mechanics adopted here, 
some events are not determined by their past histories.
The question of foreknowledge and free will can therefore be reexamined in the light of the 
modern physical understanding of causal order and physical determination.

The intuitive connection between foreknowledge and determination may come partly from an assumption 
about \emph{how} knowledge of the future must be acquired,
viz.,
that such knowledge must ultimately be grounded in antecedent conditions sufficient to determine 
the known event.
The example described in this paper shows that this is not always the case:
epistemic access to an event need not track its causal determination.

\subsection*{AI tools}
Large language model and AI-assisted tools, including ChatGPT (OpenAI), GitHub Copilot, 
and Google Search AI Mode, were used during manuscript preparation for language editing,
assistance with manuscript organization, literature searches,
and critical discussion of arguments and presentation.
I examined all suggestions, references, and sources, 
and I take full responsibility for the content of the manuscript.

\bibliography{references}

\end{document}